\RequirePackage{fixltx2e}
\documentclass[10pt,journal]{article}
\usepackage[latin9]{inputenc}
\usepackage{geometry}
\usepackage{amsmath}
\usepackage{amsthm}
\usepackage{amssymb}
\usepackage{graphicx}

\theoremstyle{plain}
\newtheorem{thm}{\protect\theoremname}
\theoremstyle{plain}
\newtheorem{lem}[thm]{\protect\lemmaname}
\theoremstyle{definition}
\newtheorem{defn}[thm]{\protect\definitionname}

\usepackage{amsthm}\interdisplaylinepenalty=2500
\usepackage[rightcaption]{sidecap}
\usepackage{fixltx2e}
\usepackage{dblfloatfix}

\usepackage{dblfloatfix}
\usepackage{url}
\usepackage{wrapfig}
\usepackage{lscape}
\usepackage{rotating}
\usepackage{epstopdf}
\usepackage[nottoc]{tocbibind}

\title{Information Theory and Direction Selectivity}

\author{Aman Chawla}

\providecommand{\definitionname}{Definition}
\providecommand{\lemmaname}{Lemma}
\providecommand{\theoremname}{Theorem}

\begin{document}
\title{Information Theory and Direction Selectivity}
\author{Aman Chawla}
\maketitle
\begin{abstract}
In this brief paper, the authors study the tuning curves of starburst
amacrine cells (SACs) and introduce a quantity called the irresolution
or ambiguity of a SAC. They show that the rate of data generated by
a starburst amacrine cell is inversely proportional to its irresolution.
This is done by providing bounds on the rate required to encode the
generated data. This technique can be applied to different cell types
with different tuning curves. In this manner, information theoretic
views can be introduced to the cases of biological cells which are
not normally considered as transmitters of information as say retinal
ganglion cells are. The intuition that even such cells generate information
is thus quantified. 
\end{abstract}
The detection of motion is a need of living organisms. In the wakeful
state of consciousness, things are not necessarily still. Motion may
only be detected against a background stillness. Therefore to identify
motion, more than a single cell may be needed. In the retina, this
is therefore one of the earliest places where co-operation and co-ordination
between cells is needed.

\begin{figure*}
\centering \includegraphics[width=18cm]{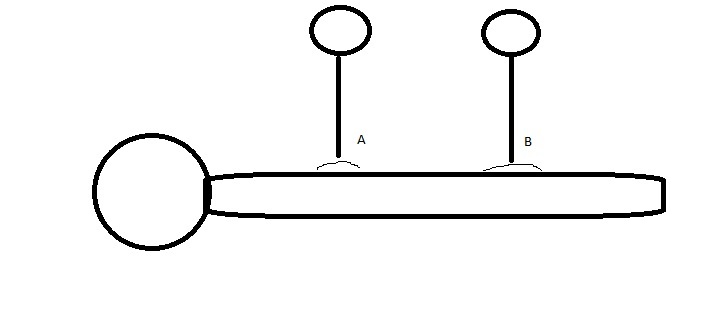} \caption{Reichardt detector}
\label{fig: reich} 
\end{figure*}

\begin{figure*}
\centering \includegraphics[width=18cm]{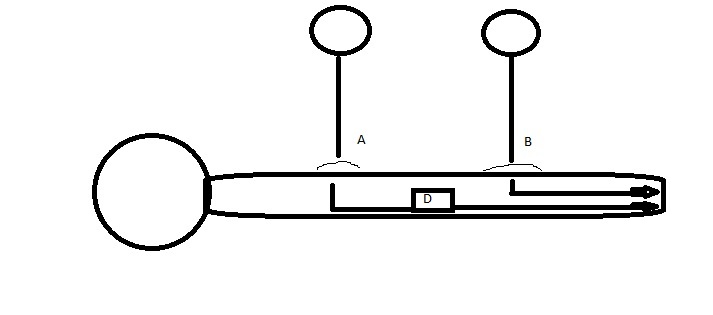} \caption{Rall mechanism}
\label{fig: rall} 
\end{figure*}

There are two main known mechanisms for direction selectivity: The
Reichardt detector, shown conceptually in Figure \ref{fig: reich}
and the Rall mechanism, Figure \ref{fig: rall}.

The Reichardt detector works as follows. The bipolar cells A and B
impinge on the dendrite under consideration at locations A and B,
respectively. The input of bipolar cell A is broad, while that of
bipolar cell B is narrower. Suppose the stimulus moves from left to
right. Then, first the broad peak is stimulated and then the narrower
peak. When these peaks reach the terminal, they get summed. The narrower
then gets raised by the broad and they together cross the threshold.
Suppose the stimulus moves from right to left. Then, first the narrower
peak is stimulated and then the broader one. As a result, there is
little or no overlap between the two peaks, and even if summed, they
will not cross the threshold. This is what gives rise to direction
selectivity.

In the Rall mechanism there is a delay element, D, introduced by the
axoplasm between A and B. This delay causes the signals from A to
be broadened when reaching the terminal, whereas the signals from
B remain narrow. A similar summation effect as the Reichardt mechanism
then leads to directional preference.

The Rall and the Reichardt mechanisms have been known for some time.
Several experiments have tried to discriminate between the two mechanisms,
to pick the correct one. The predicted observations for the Reichardt
mechanism are found to match more closely the experimental results
than those of the Rall mechanism. 

\begin{figure}
\includegraphics[width=5in]{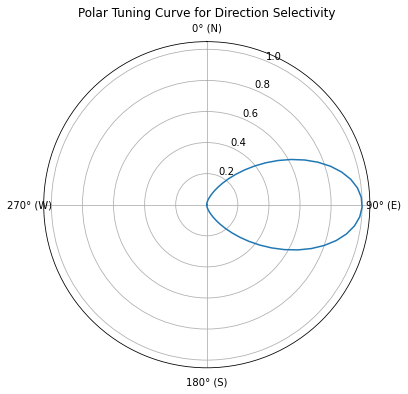}

\caption{Polar Tuning Curve. \label{fig:Polar-Tuning-Curve.}}

\end{figure}

\begin{figure}
\includegraphics[width=5in]{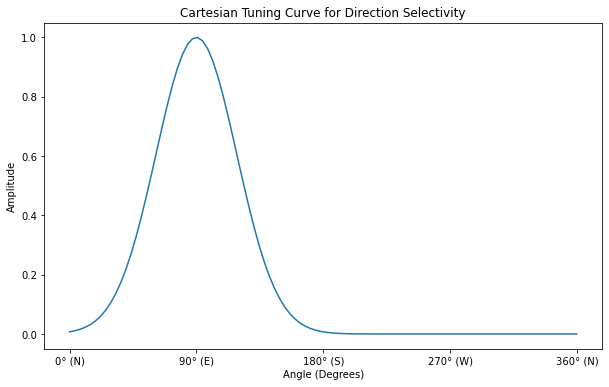}

\caption{Cartesian version of polar tuning curve. \label{fig:Cartesian-version-of}}

\end{figure}

From the theorists perspective, can we consider the dendrite as a
channel? Alternately, it is a source encoder. The question is, what
is it that qualifies an encoder as being good. And if a dendrite is
a communication channel, what is the capacity of the channel? Does
there lurk a rate-distortion problem? From the experimenter's perspective,
the stimulus is presented on the screen and an output is obtained
from the apex of the dendrite. So it does seem to be a communication
channel. Figures \ref{fig:Polar-Tuning-Curve.} and \ref{fig:Cartesian-version-of}
show how the experimentalist observes the output.

\begin{figure*}
\centering \includegraphics[width=18cm]{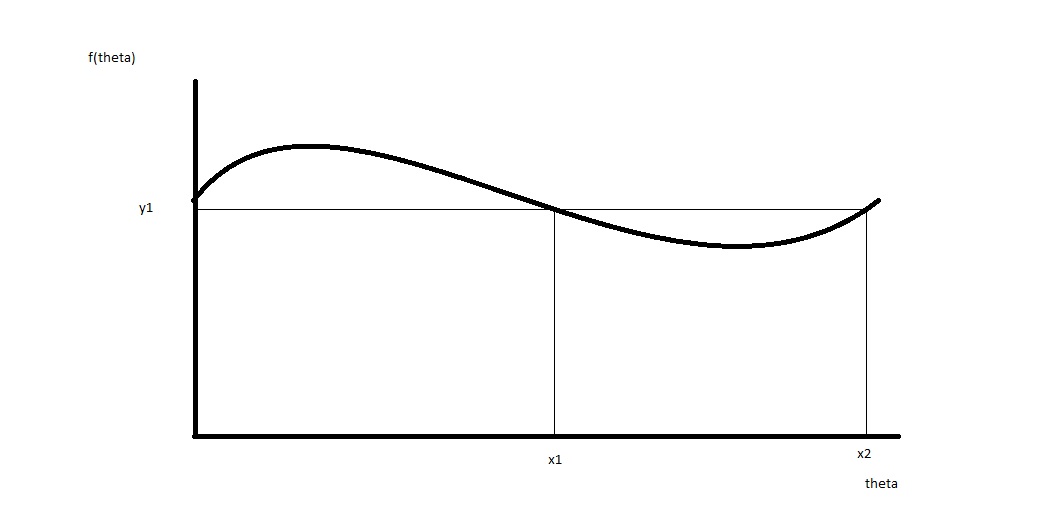} \caption{The response of a starburst amacrine cell to different input directions,
depicted as an arbitrary function of angle theta. The image y1 corresponds
to two preimages x1 and x2.}
\label{fig: axonfournodese3} 
\end{figure*}

Consider Figure \ref{fig: axonfournodese3}. It shows the response
of a starburst amacrine cell to stimulation presented at various input
directions. The response is given by the curve $f(\theta)$. 

Suppose the experimenter presents an angle of $\theta_{1}$ degrees.
The cell has a response curve $f_{1}(\theta)$ so it produces an amplitude
response $f_{1}(\theta_{1})$. This response is corrupted by noise
$\eta_{1}$ so that the final output with the experimenter is 
\begin{equation}
y_{1}=f_{1}(\theta_{1})+\eta_{1}.\label{eq: 1}
\end{equation}
It is the task of the cell, we can assume, to allow the recovery of
the direction (in degrees) from $y_{1}$. Here we show that it is
impossible to recover the angle uniquely.
\begin{lem}
Recovery Lemma. Given a cellular response $y_{1}$ (see Eqn. \ref{eq: 1}),
it is impossible to recover the directional angle in degrees uniquely,
for an arbitrary response function $f_{1}(\cdot)$. 
\end{lem}

\begin{proof}
To prove this lemma we need to come up with one example of a response
function where it is not possible to recover the directional angle
uniquely. One such example is shown in Figure \ref{fig: axonfournodese3}.$\qed$
\end{proof}
Next we quantify the error in recovery. 
\begin{thm}
\label{th: 1} Given a cellular response $y_{1}$ (see Eqn. \ref{eq: 1})
and a response function $f_{1}(\cdot)$, the error (representational)
in recovering the input direction is bounded from above by $\log_{2}(k)$
bits where $k$ is the maximum number of angles corresponding to a
value of the response function. From rudimentary graphical considerations,
$k=2*n-1$ where $n$ is the number of critical points (maxima or
minima) of the (continuous) response function. 
\end{thm}

\begin{proof}
Suppose we are using $l$ bits of precision to represent a real number.
The cellular response must be used to first estimate the value of
the response function $f_{1}(\cdot)$ at the transmitted angle $\theta$,
in the presence of Gaussian noise. The error analysis for recovering
the value of the response function, would be the same as that for
an additive white Gaussian noise channel, which has a probability
of error $Poe$ which decays with block length $N$, given by the
reliability function: 
\begin{equation}
Poe\approx\exp^{-Ng(R,N)}
\end{equation}
where 
\begin{equation}
g(R,N)=C/2-R
\end{equation}
for low rates and 
\begin{equation}
g(R,N)=(\sqrt{C}-\sqrt{R})^{2}
\end{equation}
for high rates less than capacity \cite{gallager1968information}.
Under this probability of error decay law, the value of the response
function can be estimated to within 
\begin{equation}
POE\leq\exp^{-lg(R,l)}\label{eq: poe}
\end{equation}
at the data rate $R$ bits per second. This means that $l\times R$
bits are transmitted in $l$ seconds.
\end{proof}
We have next to recover from the value of the response function, the
actual angle that was transmitted and we immediately see that if there
are at most $k$ angles corresponding to the estimated value of the
response function, then $\log_{2}(k)$ bits would suffice to recover
the precise angle. 
\begin{defn}
Irresolution
\end{defn}

As per Theorem \ref{th: 1} a total of $s$ bits must be transmitted
by the cell for every $l$ seconds of directional information transmission,
where $l\times R\leq s\leq l\times R+\log_{2}(k)$. We send $c=s/(l*R)$
bits for every input bit. $c$ is thus dimensionless. Clearly, $1\leq c\leq1+\frac{\log_{2}(k)}{l*R}$.
Since this allows error free recovery, this number is akin to a performance
metric for the system. We can call it the \textbf{irresolution} of
the cell. Clearly, if irresolution increases for a change in coding
mechanism (Rall, Reichardt, etc.), the new mechanism is taking fewer
bits in for every output bit and so is less effective in resolving
the input angle. This allows us to quantify potential coding mechanisms'
performance by simply looking at the response function. $\qed$

\begin{figure*}
\centering \includegraphics[width=18cm]{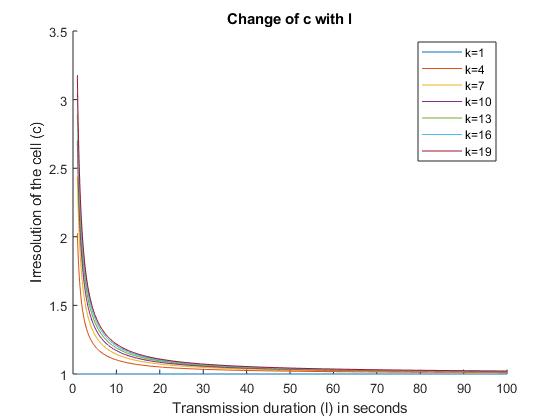} \caption{Variation in the irresolution of a cell with increasing duration of
transmission and varying directional ambiguity $k$}
\label{fig: axonfournodese4} 
\end{figure*}

This definition is explored in Figure \ref{fig: axonfournodese4}.
The irresolution goes down as the transmission duration increases.
The curves look like rate-distortion functions. We will explore this
parallel shortly.

\begin{figure*}
\centering \includegraphics[width=6in]{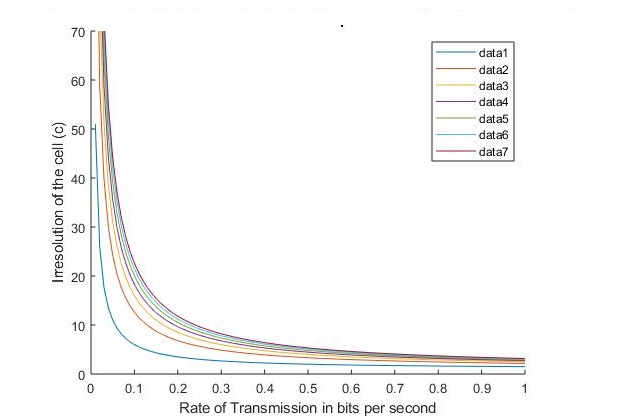}\caption{Irresolution-rate formulation: Bound on the irresolution of a cell
with increasing rate of transmission (and a total transmission duration
of 2 seconds) and varying directional ambiguity $k$. In the legend,
$k$ goes from 2 through 20 in steps of 3.}
\label{fig: axonfournodese5} 
\end{figure*}

We can ask the question, which coding scheme is optimal for transmitting
in the presence of irresolution. We can also allow different mechanisms
to send the image and the preimage resolving bits and compare the
resulting achieved bit rates with a biophysically plausible cellular
mechanism such as the Rall or Reichardt mechanisms discussed above.
The irresolution of the cell is akin to a distortion and we can plot,
on the horizontal-axis, the rate instead of the duration of transmission.
This is shown in Fig. \ref{fig: axonfournodese5}.

To conclude, in this paper we introduced a simple information theoretic
analysis of the tuning curve generated by a starburst amacrine cell.
We introduced a new quantity called the irresolution of a cell and
provided illustrations of how it varies with directional ambiguity,
a parameter we defined: $k$.

 \bibliographystyle{unsrt}
\bibliography{directionSelInfov8}
 
\end{document}